 \documentclass[preprint2]{aastex}

\usepackage{xspace}
\usepackage{pstricks}
\newsavebox{\IBox}
\usepackage{yhmath}
\usepackage{tikz}
\usetikzlibrary{shapes,arrows}

\usepackage{hyperref}
\hypersetup{
    bookmarks=true,         
    unicode=false,          
    pdftoolbar=true,        
    pdfmenubar=true,        
    pdffitwindow=false,     
    pdftitle={On the reliability of microvariability tests in quasars},    
    pdfauthor={Dr. J.A. de Diego},     
    pdfsubject={},   
    pdfcreator={},   
    pdfproducer={},  
    pdfkeywords={} {} {}, 
    pdfnewwindow=true,      
    colorlinks=false,       
    linkcolor=red,          
    citecolor=green,        
    filecolor=magenta,      
    urlcolor=cyan           
}

\newcommand{\myemail}{jdo@astro.unam.mx}
\def \chisq {\ensuremath{\chi^2}\xspace}
\def \ftest {$F$-test\xspace}
\def \ctest {$C$-test\xspace}
\def \aov {ANOVA\xspace}
\def \rw {$N(0,0.006)$ RW\xspace}
\defcitealias{diego:2010}{Paper~I}
\defcitealias{gaur:2012}{G2012}

\slugcomment{Submitted to The Astronomical Journal}

\shortauthors{de Diego}

\begin{document}






\title{On the reliability of microvariability tests in quasars}


\author{Jos\'e A.\ de Diego\altaffilmark{1,2}}
\affil{$^1$ Instituto de Astronom\'{\i}a, 
	Universidad Nacional Aut\'onoma de M\'exico, 
    Avenida Universidad 3000,
    Ciudad Universitaria,
    C.P. 04510, Distrito Federal, Mexico}
\affil{$^2$ Instituto de Astrof\'{\i}sica de Canarias - Universidad de La Laguna, CEI Canarias: Campus Atl\'antico Tricontinental, E38205 - La Laguna, Tenerife, Spain}
\email{\myemail}




%

\begin{abstract}
Microvariations probe the physics and internal structure of quasars. 
Unpredictability and small flux variations make this phenomenon elusive and difficult to detect. %
Variance based probes such as the $C$ and $F$ tests, or a combination of both, are popular methods to compare the light-curves of the quasar and a comparison star.  
Recently, detection claims in some studies depend on the agreement of the results of the $C$ and $F$ tests, or of two instances of the \ftest, in rejecting the non-variation null hypothesis. 
However, the \ctest is a non-reliable statistical procedure, the \ftest is not robust, and the combination of tests with concurrent results is anything but a straightforward methodology. 
\emph{A priori} Power Analysis calculations and \emph{post hoc} analysis of Monte-Carlo simulations show excellent agreement for the Analysis of Variance test to detect microvariations, as well as the limitations of the \ftest. 
Additionally, combined tests yield correlated probabilities that make the assessment of statistical significance unworkable. 
However, it is possible to include data from several field stars to enhance the power in a single \ftest, increasing the reliability of the statistical analysis.
This would be the preferred methodology when several comparison stars are available. 
An example using two stars and the enhanced \ftest is presented.
These results show the importance of using adequate methodologies, and avoid inappropriate procedures that can jeopardize microvariability detections. 
Power analysis and Monte-Carlo simulations are useful tools for research planning, as they can reveal the robustness and reliability of different research approaches.
\end{abstract}



\section{Introduction}

%
Flux variability provides unique information about the physics and geometry of the unresolved central source in Active Galactic Nuclei (AGNs). %
%
%
Variability seems to be present in every AGN, usually exhibing increasing amplitude at shorter wavelengths and at longer time scales. 
Quasar optical light-curves generally have variations of about 10\% on timescales of months, and the power spectra of these light-curves are consistent with random walk processes \citep{peacock:1983,koen:1994,kelly:2009}.
At optical frequencies, the shortest time scale variations last from some minutes to few hours \citep[e.g.][]{kidger:1990, carini:1992, gopal:1995, jang:1997, ramirez:2004}, and have amplitudes about a few hundredths of one magnitude. 
This phenomenon is called microvariability and, due to causality arguments, it is thought to arise from an inner region of a few light-minutes in size. 

Charge Couple Device (CCD) differential photometry techniques have enhanced the accuracy of variability studies. However, microvariability reports have always been regarded suspiciously because changes in flux are comparable to photometric errors. 
Therefore, instrumental limitations, varying atmospheric conditions, observational strategy, data reduction and analysis methodology are critical issues that should be carefully handled by the researcher to produce trustworthy results. 
During the past seventeen years there has been an interest in developing highly reliable statistical procedures for detecting optical microvariations in quasar. 
Thus, \citet{jang:1997} developed the \ctest, a quite simple methodology that became a popular method to analyze quasars' light-curves. 
By the same time, the one-way analysis of variance (\aov) test was adapted by \citet{diego:1998} for microvariability studies. 
The use of \aov in light-curve studies has helped to challenge the common understanding of the quasar microvariability phenomenon by closing the gap between radio-loud and radio-quiet quasars differences \citep{ramirez:2009}. 
Moreover, \aov methodology is unique in the possibility of assessing internal estimate of errors, while providing a high test power.

Unfortunately, \aov requieres that data should be gathered in time scales much shorter than variability scales to avoid that the data dispersion inside the groups could be dominated by flux variations rather than from errors.
Therefore, performing the \aov test requires an extra effort in data gathering and reduction due to the need of somehow oversampled light-curves, as well as rather tedious calculations. 
These circumstances have limited the usage of \aov in microvariability studies, even if it is a robust and powerful statistical procedure that has been employed for decades in other related areas, and particularly in periodicity studies on folded light-curves of variable stars  \citep[e.g. ][]{schwarzenberg:1989,schwarzenberg:1996,wozniak:2004,graham:2013}. 

Lately, \citet[Paper I from now on]{diego:2010} analyzed the performance of a number of microvariability tests, and demonstrated that both the \ctest and the \chisq-test were not trustworthy, the former due to wrong design, and the later because its reliability depends on the exact quantification of all the error sources, which is impractical to say the least. 
On the contrary, the \ftest and \aov showed to be reliable tests, and \aov stood out as the most powerful test of those analyzed. 
The \ftest does not require additional efforts in comparison, for example, to the \ctest, and thus \citetalias{diego:2010} has contributed to popularize its use \citep[e.g.][]{joshi:2011, joshi:2013, paliya:2013, goyal:2013}. 
Nevertheless, interest for developing new test procedures has not decayed, and a number of test adjustments have been considered. 
Thus, \citet{joshi:2011} have proposed a modification to the \ftest to scale photometric error estimates to compare sources with different brightnesses. 
Modification to the \ctest by \citet{goyal:2013} have been proposed to correct for the original test inadequacies, although the corrected \ctest is just a square root transformed version of the \ftest \citepalias[see also][]{diego:2010}, and thus both yield the same probabilities \citep[cf.][table 2]{goyal:2013}. 

Another strategy consists in producing a battery of tests, statistically known as multiple testing or \emph{multitesting}.
Multitesting is used in several science disciplines, and particularly in genomics where it is necessary to perform many inferences to test different null hypothesis over different variable sets of the same high-dimensional multivariate data, and adjusting the probability values to avoid increasing the total number of Type I errors.
In this context, each inference is associated to a single null hypothesis, and thus the result of every test is considered apart from the results of the other tests.
To keep control on the Type~I errors, the probability of false discoveries is addressed either through familywise error rate procedures (of which the Bonferroni correction is considered the simplest and most conservative), or the less stringent false discovery rate procedures.
What makes quasar microvariability multitesting different from most multitesting applications is that it is employed in \emph{only two} tests to  achieve a \emph{single} inference probing the \emph{same} null hypothesis (non-variability) over the same \emph{univariate} photometric data.

The single inference made in quasar microvariability multitesting is to consider a variation event reliable only if \emph{both} tests agree in the rejection of the non-variability null hypothesis \citep{joshi:2011, hu:2013}. 
This approach contrasts sharply with the way that the assessment of overall significance of multiple tests probing the same null hypothesis is addressed in statistical literature. 
Based on Fisher's method of combining probabilities through the \chisq statistics, \citet{brown:1975} and \citet{kost:2002} developed covariance based tests to prove the distribution of a sample of probability values.
Unfortunately, the covariances between the involved tests should somehow be estimated from a reference sample, which is unfeasible for quasar microvariability studies of single light-curves.
Recently, \citet{vovk:2012} discussed the problem of multiple testing of a single null hypothesis in the context of multivariate probabilities and copulas, providing an example for combining two tests. 

The statistical methodology employed for a given research is an important aspect of the experimental design.
Different statistical tests have different strengths and weakness, they are more or less appropriate to be applied depending on the characteristics of data sets, and therefore they usually yield different probabilities.
Every test relies on a number of assumptions and conditions that must be met by the sample data and the parent population.
Violating these assumptions and conditions affects the test validity.
In practice, this means that the test would yield a larger number of Type~I or Type~II errors than expected, i.e. either the significance of the test or its power (or both) will be other than predicted.
Moreover, poor understanding of test capabilities and possible violation of test assumptions and conditions often jeopardizes the results obtained from an otherwise well grounded research.

However, aside from the \chisq-test, the \ftest and \aov for which there are sound backgrounds both in statistical literature and practical applications, the other inferential statistical procedures used so far in quasar microvariability studies lack of both, theoretical background and empirical studies of reliability. 
Important questions are neither addressed, nor even mentioned, such as test assumptions (parent distribution, minimum sample size, homoscedasticity), sampling conditions (randomness, independence), robustness (test performance when conditions are violated), power analysis (test performance in rejecting false null hypotheses), or systematic comparison with different test methodologies beyond a few empirical instances. %

This paper combines analytical studies, simulations, and observational results from recent literature to characterize and understand important aspects of the statistical methodologies used to detect AGN microvariations.
For this purpose, it presents an analysis of the reliability and power of three parametric tests, and two nonparametric  methodologies.
The paper also considers the effect of combined probabilities and tests correlations on multitesting, and provides a new procedure to integrate several comparison stars light-curves in the statistical analysis.
Altogether these studies show the power and reliability of the \aov and Bartels test for microvariability studies, and the necessity of developing trustworthy methods to integrate the light-curves of several comparison stars in the analysis.

This paper is organized as follows. 
Section~\ref{sec:methodology} describes the methodology used to evaluate the analytical tests and perform simulations.
Section~\ref{sec:analytical} presents the analytical power study for \aov and the \ftest.
Sections~\ref{sec:empirical} and \ref{sec:multitesting} describe the results obtained using different statistical procedures and simulated data.
Section~\ref{sec:observations} compares the previous analysis and results extracted from the literature.
Finally, Section~\ref{sec:conclusions} presents the conclusions.

\section{Methodology}\label{sec:methodology}

This study involves power analysis calculations and data simulations to compare the performance to detect microvariability of three parametric tests (\ctest, \ftest and \aov), and two nonparametric tests (Runs test and Bartels test). 
Besides, we will address the problem of correlations between different tests and how they affect the multitest results. 
The general statistical procedure used for power analysis calculations involving the $F$ statistics (\ftest and \aov) is through noncentral $F$ distributions (see Appendix~\ref{app:ncd}). 
Statistical tests and power analysis computations have been performed using R code \citep{r:2013}.


\begin{figure*}[t]
\centering
Fig. (a) in page 24, (b) in page 25
\caption{
    Power at the significance level of $\alpha = 0.001$ for \aov with 7 groups (solid lines) and the \ftest (dotted lines). 
    Panel (a): For a variation of amplitude 0.04~mag, \aov power is larger than the \ftest power for a total number of observations of $N \geq 21$.
    Panel (b): Compared to the previous example, \aov has relatively lower power for variations of amplitude 0.02~mag, but the \ftest is practically insensible.
    See details in the text.
    }\label{fig:power}
\end{figure*}

Most differential light-curve simulations for variable quasars have been carried on using a random walk model with Gaussian drifts of different amplitudes. %
As in \cite{kelly:2009}, light-curves were generated in magnitudes rather than in fluxes, but in our case non-damped random walks were performed for variable quasars. %
Damped random walks are useful to study long-timescales variability, where a \emph{base} state can be identified for every quasar; however, in this short-timescale study this constraint may be disregarded. 
Every light-curve consists of a multiple of 5 number of observations, between 15 and 35, to enable \aov tests on homogeneous groups of 5 elements. 
Random walk Gaussian drifts $s_i$ are normally distributed with mean 0 and standard deviation $\sigma$ [$s_i \sim N(0,\sigma)$], with $\sigma = 0.006$~mag for a variable quasar.
The $\sigma = 0.006$~mag value was chosen because it yields variations that are neither too large nor too small to compare the performance of the different tests.
For this model, the last point of $N$ random walks will be distributed with a mean of 0, and a standard deviation of $\sigma \sqrt{N}$.
Finally, a Gaussian random white noise with an error $e_i \sim N(0,\varepsilon)$, where $\varepsilon=0.01$~mag, has been added to each observation to account for photometric uncertainties. %
Therefore, the random walk sequence was computed by:
\begin{equation}
    \zeta_i = \left\{ \begin{array}{ll}
                 s_i & \mbox{if $i = 1$}; \\
                 \zeta_{i-1} + s_i & \mbox{if $i \geq 2$},
                 
                     \end{array}
                     \right.
\end{equation}
where $\zeta_i$ is the true magnitude of the $i$-observation of the target object. %
Then the actual simulated measurement $z_i$ was obtained by:
\begin{equation}
 z_i = \zeta_i + e_i.
\end{equation}
In the case of non-variable objects, observations $m_i$ in every light-curve are randomly Gaussian distributed with mean zero and standard deviation given by the photometric error $\varepsilon = 0.01$~mag [$m_i \sim N(0,\varepsilon)$], except for only one simulation which error was set at 0.0181 (see below).

\begin{figure*}[t]
\centering
Fig. (a) in page 26, (b) in page 27
\caption{
    Simulated light-curves for a variable quasar (open circles) and a 
    non-variable star (asterisks) extracted from the simulated set of step modeled microvariations. 
    Light-curves comprise 35 individual data points for the quasar and the star.
    Quasar variations correspond to a segment of 5 contiguous observations, with
    an amplitude of 0.04~mag.
    Panel (a): The light-curves.
    Panel (b): The same light-curves binned in groups of 5 observations with the error 
    bars obtained from the standard error of each group, as in the \aov methodology.
    }\label{fig:group_lightcurves}
\end{figure*}

This work includes five different runs of simulations.
The first run is for 1000 light-curves of 35 observations each, with variations modeled by a step function; these simulations were compared with analytical powers calculated for both, \aov and \ftest; this run also includes 1000 extra light-curve simulations with normal distributed data and an error of $\varepsilon_r = 0.0181$ that was used to compare non-Gaussianity in the difference between the analytic and simulatd powers for the \ftest .
The second run is for 3000 light-curves, 600 for each of the 5 groups of $n =$ 15, 20, 25, 30 and 35 observations, with variations modeled as random walks; these simulations were used to study the performance of each individual test.
The third run is for 30,000 non-variable light-curve simulations with $n = 35$ observations, that are used to check the number of Type~I errors.
The fourth run is for 300 light-curves of $n =$ 15, 20, 25, 30 and 35 observations, with variations modeled as random walks; these simulations were used to compare the relationships between the different tests, while producing graphs that are not too messy to make data patterns eye-catching.
The last run consists of 3000 variable and 3000 non-variable light-curves of $n = 35$ observations, the variable curves modeled as random walks; these simulations were performed to show how to combine two or more light-curves obtained from different comparison stars to improve the power of the \ftest. 

Finally, some of the blazar microvariability results in \citet[G2012 from now on]{gaur:2012} have been analyzed to compare with the simulations. 
These authors have studied blazar light-curves using several of the tests mentioned above, and thus it is an excellent example to prove some of the methodologies presented here.

\section{Comparing analytical and empirical powers for \aov and \ftest}\label{sec:analytical}

It is interesting to study the power of a test from both mathematical analysis (when possible) and Monte Carlo simulations, and to compare both results to evaluate the effect of violation of test conditions and assumptions on the test reliability.
Moreover, the comparison of such study between different tests is very helpful to indicate the most appropriate methodology to analyze the data.
Analytical power can be studied only on parametric tests, such as \ftest and \aov.
Therefore, for this analytical study we are considering neither Bartels nor Runs tests, as they are non-parametric.
Study of the power for non-parametric tests can be performed only through Monte Carlo simulations \citep[e.g.][]{tanizaki:1997,mumby:2002}.
With respecto to the \ctest, it is not a reliable statistical procedure \citepalias{diego:2010}, and therefore its power must be characterized empirically rather than analytically.
Accordingly, we will address the Bartels, Runs, and $C$ test power issues later, when examining simulations.

In general, the power of a test depends on three parameters: the level of significance, the sample size, and the effect size.
In our case, the sample size is related to the number of observations included in light-curves, and the effect size is a measure of the amplitude of the variations with respect to the photometric errors.

To perform the power analysis of the \ftest and \aov, consider $N$ observations with an error of 0.01~mag, and a step function for variations of 0.04~mag that affect $N/7$ of contiguous data. 
For the \ftest, consider $N=2, 3, 4\ldots 100$ (in this case it is not necessary for $N/7$ to be an integer number), and the effect size (see Appendix~\ref{app:ncd}) will be $r=0.04^2/(7*0.01^2) = 2.2857$.
In the case of \aov, the data is divided in $k=7$ groups of the same number $n = 2, 3, 4\ldots 15$ observations (i.e. a \emph{balanced} \aov), and a total number of $N = 7 n$ data points. 
In this particular case, the \aov effect size (see Appendix~\ref{app:ncd}) may be easily calculated as the square root of the \ftest effect size ($f = \sqrt{r} = 1.5119$).

Figure~\ref{fig:power} presents a comparison of the power of \aov and \ftest as a function of the number of observations $N$, at the level of significance of the tests of $\alpha = 0.001$. 
The \ftest power is approximated by a continuous curve as the number of observations can be any positive integer number, while the \aov power has been calculated for 7 groups, and thus it is displayed as a stair plot with step jumps at multiples of 7 observations.
Figure~\ref{fig:power}a shows the power for variations of amplitude 0.04~mag. In this case, it is obvious that the power of \aov is larger than the power of the \ftest when $N \geq 21$ (i.e. 3 observations per group).
For $N = 35$ (5 observations per group), \aov almost gets the maximum power (0.998), while the for the \ftest it is 0.610.
As the test power also depends on the effect size, if the variations are large enough any test will detect them. 
Thus, the importance of choosing the most appropriate test is stressed for small effect sizes.
Figure~\ref{fig:power}b shows the power of both \aov and \ftest when the variations have amplitudes of only 0.02~mag (i.e. effect sizes of $r = 0.5714$ for the \ftest and $f = 0.7559$ for \aov).
As this figure shows, the \ftest is practically insensible to such small variations, while the \aov still keeps some power at small number of observations.

\begin{figure}[t]
\centering
Fig. in page 28
\caption{
    Quasar light-curves may show non-Gaussian profiles.
    This figure shows an example of a possible distribution, based on simulated
    light-curves for a variable quasar.
    The histogram of magnitudes for the 35 observations of the light-curve shown in Fig.~\ref{fig:group_lightcurves}b (solid line), and the scaled probability density function for the simulated set of step modeled light-curve microvariations (dashed line).
    }\label{fig:gaussianity}
\end{figure}

Let us compare the analytical results shown above with powers obtained from simulations.
We concentrate in the particular case of a total of $N = 35$ observations, performing 1000 light-curve simulations.
Figure~\ref{fig:group_lightcurves} shows an example of such light-curve simulations. 
The step function like variation of the quasar shows up around 18 time units, and it presents an increase in brightness of 0.04~mag (from 17 to 16.96~mag).
The empirical power is calculated as the ratio of the number of detections and the number of simulations.
The \aov power obtained from these simulations is $0.998 \pm 0.001$, in excellent agreement with the analytical power (0.998).
Howeveer, the power for the \ftest yielded by the same simulated light-curves is $0.53 \pm 0.02$, i.e. about a 13\% less than the analytical power (0.610).

\begin{figure*}[t]
\centering
Fig. (a) in page 29, (b) in page 30
\caption{Simulated light-curves for a non-variable star (left) and a quasar (right).
    The number of data points in the light-curves range from 15 to 35, as in this 
    example.
    The random walk step distribution for the quasar light-curve is Gaussian, 
    with a mean $\mu=0$ and a standard deviation $\sigma=0.006$ (i.e., \rw).}
    \label{fig:lightcurves}
\end{figure*}

It is worth to investigate this difference between the analytical power of the \ftest and the power calculated from simulations because it may give us clues about the \ftest limitations.
Figure~\ref{fig:gaussianity} shows the probability density distribution (PDF) of the simulations, and the histogram for the magnitudes corresponding to the same quasar light-curve shown in Figure~\ref{fig:group_lightcurves}a.
The PDF is the result of two Gaussian profiles with standard deviation $\sigma=0.01$ mag, but one with a mean of 17~mag and the other with a mean of 16.96~mag.
The regions below the respective PDF curves are 6/7 and 1/7 of the total area.
The histogram of the simulated light-curve evidences the effect of the PDF profile in the data.
Clearly, for the simple variability model considered in the simulations, the bimodal distribution shown in Figure~\ref{fig:gaussianity} is not a Gaussian.
But Gaussian distribution is an important condition to perform a reliable \ftest, which is known to be very sensible to non-normality \citep{box:1953}.
Therefore, we can attribute the difference between the analytical power and the power obtained from simulations to the non-normal distribution of the variable quasar photometric data.
To check if non-Gaussianity is responsable for this lack of agreement, we need simulations of non-variable, normal distributed light curves but with a dispersion that produces the same effect size $r$ obtained for the variable sample.
This can be done by setting the sample variance $\sigma_r^2$ to the value:
\begin{equation}
\sigma_r^2 = \varepsilon^2 (1+r), 
\end{equation}
which yields $\sigma_r^2 = 3.286 \times 10^{-4}$ (see Appendix~\ref{app:ncd}), and it is analogous to a photometric error of $\varepsilon_r = \sigma_r = 0.0181$. 
Simulating 1000 N(0, 0.0181) light-curves, and comparing them with N(0, 0.01) simulations, yields an empirical power of $0.610 \pm 0.005$ that agrees with the analytical value.
This result confirms that the lack of power agreement between the step-like variations and the analytical \ftest power is due to non-normality.

In real situations, the variability in quasar light-curves may be (and probably are) more complicated than the simulations presented here, presenting different states of variability and producing multimodal data distributions. 
For this reason, the \ftest for homogeneity of variances has a power lower than expected if the test conditions were met, and thus it is less reliable than \aov or, as we will see later, the Bartels test.

\begin{table*}[t]
\begin{center}
\caption{Power for tests at $\alpha = 0.01$ for \rw light-curves.}\label{tab:proptests}
\begin{tabular}{lr@{$\,\pm\,$}lr@{$\,\pm\,$}lr@{$\,\pm\,$}lr@{$\,\pm\,$}lr@{$\,\pm\,$}lr@{$\,\pm\,$}lr@{$\,\pm\,$}lr@{$\,\pm\,$}lr@{$\,\pm\,$}lr@{$\,\pm\,$}}
 \tableline\tableline 
N & \multicolumn{2}{c}{\ftest} & \multicolumn{2}{c}{\aov} & \multicolumn{2}{c}{Bartels} & \multicolumn{2}{c}{Runs} & \multicolumn{2}{c}{\ctest} \\ 
 \tableline 
  15 &        0.11 &        0.01 &        0.22 &        0.02 &        0.18 &        0.02 &        0.09 &        0.01 &       0.007 &       0.003 \\ 

  20 &        0.26 &        0.02 &        0.39 &        0.02 &        0.35 &        0.02 &        0.18 &        0.02 &       0.015 &       0.005 \\ 

  25 &        0.39 &        0.02 &        0.50 &        0.02 &        0.49 &        0.02 &        0.26 &        0.02 &       0.027 &       0.007 \\ 

  30 &        0.53 &        0.02 &        0.63 &        0.02 &        0.60 &        0.02 &        0.36 &        0.02 &        0.06 &        0.01 \\ 

  35 &        0.64 &        0.02 &        0.74 &        0.02 &        0.73 &        0.02 &        0.47 &        0.02 &        0.08 &        0.01 \\ 

\end{tabular}
\end{center}
\end{table*}

\section{More empirical power analysis}\label{sec:empirical}

In this section we will continue investigating the power of different statistical tests, including the \ctest and non-parametric tests.
Instead of the simple step function model for variability, which was adequate to stress the differences between the \ftest and \aov, the light-curve simulations employed here are based on the random walk model introduced in Section~\ref{sec:methodology}.

Figure~\ref{fig:lightcurves} shows an example of the simulated light-curves for the comparison star and the variable quasar.
The quasar light-curve for this figure was modeled using a Gaussian random walk with mean $\mu=0$ and drift $\sigma = 0.006$; we will use the compact notation $N(\mu,\sigma)$ RW [\rw in our case] to denote the random walk parameters.


Table~\ref{tab:proptests} shows the proportions of microvariability detections for different tests at the significance level of $\alpha = 0.01$, based on a total of 3000 \rw simulations divided in 5 sets of 600 simulations according to the number of observations included in the light-curves.
Column 1 shows the number of observations, namely 15, 20, 25, 30 and 35 observations per light-curve.
Columns 2 shows the proportion of detections and its error for the \ftest for a given number of observations (for example, the \ftest detects microvariability in a proportion of 0.11 of the simulations with 15 observations) .
The rest of columns are similar to columns 2, but for the different tests.
From this table, it is obvious that \aov and Bartels' test produce the largest number of detections with comparable results. 
The number of detections and the power of the \ftest is well below \aov and Bartels.
The power of the Runs test is very low, and thus this test is very limited to be used for microvariability detections.
The \ctest has the lowest power, in accordance with the results discussed in \citetalias{diego:2010} for simulations based on different light-curve variability models (Gaussian shaped and constant trend variations).
As expected, the power of the tests increases with the number $N$ of observations.
The \aov and Bartels' test have the largest powers (between 18\% and 74\%) for any number of observations, followed in decreasing order by the \ftest, the Runs test and the \ctest, the later with a minimal power ranging between 0.7\% and 8\%.


Table~\ref{tab:testcomp} shows in detail the result of the 600 \rw simulations of light-curves with $N=35$ observations each.
The first column identifies the test used for the analysis.
The second and third columns show the number of detections at the significance levels of $\alpha = 0.001$ and 0.01, respectively.
The fourth and fifth columns display the empirical power calculated from the ratio between the number of detections and the total number of simulations.
As expected, the dependence of the power with the level of significance of the test is clearly seen comparing the number of detections or the power for a given test at $\alpha = 0.001$ and $\alpha = 0.01$; the larger the level of significance, the larger the number of detections and the power.
Columns five and six show the \emph{likelihoods}, and columns seven and eight the \emph{false discovery rates} of the tests, as described below.

\begin{table*}[t]
\footnotesize
\begin{center}
\caption{Statistics for 600 \rw light-curve simulations of $N=35$ observations.\label{tab:testcomp}}
\begin{tabular}{lrrcr@{$\,\pm\,$}lr@{$\,\pm\,$}lcrrcrr}
\tableline\tableline
&        \multicolumn{2}{c}{Detections}          & &            \multicolumn{4}{c}{Power}    & &    \multicolumn{2}{c}{Likelihood}    & &    \multicolumn{2}{c}{\emph{FDR}} \\
\cline{2-3}
\cline{5-8}
\cline{10-11}
\cline{13-14}
& $\alpha = 0.001$ & $\alpha = 0.01$ & & \multicolumn{2}{c}{$\alpha = 0.001$} & \multicolumn{2}{c}{$\alpha = 0.01$} & & $\alpha = 0.001$ & $\alpha = 0.01$ & & $\alpha = 0.001$ & $\alpha = 0.01$  \\
\tableline
F-test        & 256    & 383    &    & 0.43    & 0.02    & 0.64    & 0.02     &    & 430    &    64    &    &    0.0023    & 0.0154 \\
ANOVA         & 349    & 445    &    & 0.58    & 0.02    & 0.74    & 0.02     &    & 580    &    74    &    &    0.0017    & 0.0133 \\
Bartels       & 342    & 440    &    & 0.57    & 0.02    & 0.73    & 0.02     &    & 570    &    73    &    &    0.0018    & 0.0135 \\
Runs          & 162    & 282    &    & 0.27    & 0.02    & 0.47    & 0.02     &    & 270    &    47    &    &    0.0037    & 0.0208 \\
C-test        &  11    &  46    &    & 0.018   & 0.006   & 0.08    & 0.01     &    &  18    &     8    &    &    0.0526    & 0.1111 \\
\tableline
\end{tabular}
\end{center}
\end{table*}

\subsection{Type I errors}

To check how the tests comply with the nominal level of significance, simulations of non-variable light-curves were analyzed.
For a fair test, the proportion of Type I errors yields the actual significance of the test.
Table~\ref{tab:testcompfornonvar} shows the number of Type I errors for 30,000 simulations of light-curves with $N=35$ observations each, at the levels  $\alpha = 0.001$ and 0.01 for each test.
The corresponding intervals of confidence for the number of Type I errors are also shown.
\aov, Bartels test, and the \ftest Type I errors met expected values for both significance levels.
The Runs test tends to yield few Type I errors, while the \ctest is too insensitive to produce any, evidencing  that they are not adequate to study microvariability.

\subsection{Test powers and detection reliability}

The relatively low power of the \ftest does not only compromise the number of detections, but also the likelihood of those detections.
This means, for example, that the probability of rejecting the null hypothesis of non-variability when it is false (i.e. detecting variability among those truly variable light-curves), compared to the probability of Type~I errors, is lower for the \ftest than for \aov  or Bartels test.

Let us define this likelihood $\mathcal{L}$ as the ratio between the power ($1 - \beta$) and the level of significance $\alpha$ for the test\footnote{In statistical terminology, the name for this ratio is \emph{likelihood ratio for positive tests}, and it is used in Bayesian statistics to obtain posterior probabilities.}: 
\begin{equation}\label{eq:likelihood}
    \mathcal{L} \simeq \frac{ 1 - \beta}{\alpha}.
\end{equation}
In the context of this study, the likelihood indicates how the test result is related with variations; if $\mathcal{L}>1$ the test is useful to detect variations, and if $\mathcal{L}<1$ the result is associated with the absence of variability.
As $\mathcal{L}$ is further away from 1 in either direction, the most confident we can be that the results indicate the presence, or absence, of variations.

Another quantity of interest is the false discovery rate ($FDR$), that is the probability of a Type~I error among those tests that reject the null hypothesis.
The $FDR$ is expressed as:
\begin{equation}\label{eq:fdr}
    FDR = \frac{\alpha}{\alpha + 1 - \beta}
\end{equation}

Table~\ref{tab:testcomp} shows the likelihoods and the $FDR$s for the 600 \rw light-curve simulations of 35 observations.
At the level of significance of $\alpha = 0.001$, equations (\ref{eq:likelihood}) and (\ref{eq:fdr}) yield $\mathcal{L} = 430$ and $FDR = 0.0023$ for the \ftest (see Table~\ref{tab:testcomp}).
For \aov, the results are $\mathcal{L} = 580$ and $FDR = 0.0017$,  and for the Bartels test $\mathcal{L} = 570$ and $FDR = 0.0018$.
Their likelihoods and $FDR$s indicate that these three tests, and even the Runs test ($\mathcal{L} = 270$ and $FDR = 0.0037$), are very reliable in the sense that if a variation is detected, it is almost certainly true.
However, the \ctest yields a much lower reliability ($\mathcal{L} = 18$ and $FDR = 0.0526$), and at $\alpha = 0.01$ (the level at which it is generally performed), the $FDR = 0.1111$ value indicates a fraction over 10\% of false variability detections (note that this numbers are only indicative, and their values depend on the parameters of the simulations).

In brief, it is worth to stress that using a low power test not only yields a lower number of detections, but that it also increases the fraction of false detections.


\section{Multitesting}\label{sec:multitesting}





In the recent years, multitesting has become popular as a procedure to guarantee the reliability of detections for microvariability events.
The way that multitesting is implemented is that for being considered a trusty detection, a microvariation should be discerned by \emph{two} tests at a significance level $\alpha = 0.01$ \citep[e.g.][]{joshi:2011, hu:2013, chand:2014}.

This section introduces some concepts to deal with the problem of joint probabilities and probability correlations.
The number of simulations to study test correlations has been reduced to 300 to avoid messing the figures with too many data points, while conserving enough power to make accurate inferences.
Each simulation includes a number of 15, 20, 25, 30, or 35 observations, ascribed randomly.
Most of the discussion will be centered on the results for two tests (\ftest and \aov), but it will be easily generalized to any set of two or more tests.

Below we will see that different tests applied to the same data sample and sharing the same null hypothesis (non-variability) yield strong correlated probabilities.
As a result, these correlations make difficult to interpret the actual probability associated to a variable event, and they make multitesting an inaccurate and restrictive tool for improving the quality of detections, at least in the current way that it is implemented in microvariability studies.
Despite these considerations, it is possible to improve the power of a test considering two or more comparison stars, and a simple procedure is presented for the \ftest, and compared with simulations.

\begin{table}[t]
\begin{center}
\caption{Simulations of 30,000 non-variable light-curves with $N=35$ observations.\label{tab:testcompfornonvar}}
\begin{tabular}{lrr}
\tableline\tableline
              &        \multicolumn{2}{c}{Type I errors} \\
              \cline{2-3}
              &    $\alpha = 0.001$    &    $\alpha = 0.01$ \\
\tableline
I.C. 95\%     &    19--41              &    266--334 \\
C-test        &      0                 &      0 \\
F-test        &     29                 &    312 \\
Runs          &     19                 &    232 \\
Bartels       &     20                 &    273 \\
ANOVA         &     30                 &    310 \\
\tableline
\end{tabular}
\end{center}
\end{table}

\subsection{Tests correlations}

Figure~\ref{fig:probregres} shows the probability correlations between \aov and \ftest for 300 non-variable and 300 \rw variable light-curve simulations. Similarly, Figure~\ref{fig:moreregres} shows probability correlations between \aov and Bartels test, and \ctest and \ftest for variable light-curve simulations.


\begin{figure}[!ht]
\centering
Fig. (a) in page 31, (b) in page 32, (c) in page 33
\caption{\aov and \ftest probabilities. 
    Panel (a): for non varying light-curves, the 
    probabilities are not correlated as shown by the regression line (dashed). 
    Panel (b): regression for 300 \rw simulations (dashed line), 
    and the critical values $\alpha = 0.001$ 
    (solid lines).
    Panel (c): the histogram and the conditional probability
    (thick line) based on normal-like residuals.}
    \label{fig:probregres}
\end{figure}

\begin{figure*}[t]
\centering
Fig. (a) in page 34, (b) in page 35
\caption{Logarithmic probabilities relationships between tests for 300 \rw simulations. 
    Panel (a) \aov and Bartels test probabilities are highly correlated as shown by 
    the regression line (dashed). 
    Panel (b) shows the tests correlation as well as the critical values for 
    $\alpha = 0.001$ significance 
    levels for the \ftest (vertical line), and for $\alpha = 0.001$ and 
    $\alpha = 0.01$ for the \ctest (solid and dotted horizontal lines, respectively).
    Note the contrast between the data distribution in both panels; while in panel (b) 
    ANOVA and Bartels test show a linear dependence, tests $C$ and $F$ show five 
    curve families that correspond to the five sets of data sample sizes (see text).}
    \label{fig:moreregres}
\end{figure*}

To study the correlation between two tests, let us define the joint probability density function $ p(x,y) $ that specifies the probability of joint occurrence of a pair of random variables $ (x,y) $:
\begin{align}\label{eq:jointprob}
 P(y_1 \leq y \leq y_2, & \, x_1 \leq x \leq x_2) = \notag \\
     & \int_{y_1}^{y_2} \int_{x_1}^{x_2} p(x,y) \mathop{dx} \mathop{dy} . 
\end{align}
By elemental probability theory, we can write:
\begin{equation}
 p(x,y) = \mathop{p(y|x)} \mathop{p(x)} ,  
\end{equation}
where $ p(y|x) $  is the conditional probability density, and $ p(x) = \int_{-\infty}^{+\infty} p(y|x) \, dy $ is the marginal density of $ x $. The expectation of the variable $y$ given the variable $x$ is called the conditional expectation $ E(y|x) $, and it is given by:
\begin{equation}
 E(y|x) = \int_{-\infty}^{+\infty} y \mathop{p(y|x)} \mathop{dy} .
\end{equation}
$ E(y|x) $ is a function of $ x $. In our case, $ x = \log P_F $ and $ y = \log P_A $, where $ P_F $ and $ P_A $ are the probabilities obtained from \aov and \ftest. From Figure~\ref{fig:probregres}b we can see that these variables have a linear relationship:
\begin{align}\label{eq:logaf}
 E(\log P_{A} & |\log P_{F}) = \notag \\
     & -0.8 (\pm 0.1) + 0.92 (\pm 0.05) \, \log P_F .
\end{align}
The residuals $ r $ of the regression fit are approximately normally distributed with a standard deviation $ \varepsilon $:
\begin{equation}
 r \sim N(0,\varepsilon) 
\end{equation}
Now we can calculate the probability that \aov yields a probability $ P_A > 0.001 $ depending on the result of a given \ftest $P_F$:
\begin{align}\label{eq:plogaf}
 P(\log P_A & > -3 \, | \log P_F) = \notag \\
     & \int_{-3}^{0} f(\widehat{\log P_A},\varepsilon) \mathop{d(\log P_A)} ,
\end{align}
where $ \widehat{\log P_A} $ is the expected value obtained from the regression line at a given $ \log P_F $, and $f$ denotes the normal PDF.
Note that $ \varepsilon $ has been estimated from all the residuals for \ftest values $ \log P_F < -3 $ to avoid the distortion due to $ \log P_A > 0 $ truncated values. 
This procedure introduces some analytical underestimate of the proportion of rejected \aov for small values of $ \log P_F $ between the analytical estimates from equation~(\ref{eq:plogaf} and the actual proportion obtained from the simulations.
Taken this into account results in $ \hat{\varepsilon} = 2.0125 $. Figure~\ref{fig:probregres}c shows both the analytical and the empirical probabilities.

The parameters in equation~(\ref{eq:logaf}) depend on the parent population from which the data set has been extracted. 
In the case shown above, the parent distribution corresponds to the \rw model, and thus it is easy to generate random samples to study the joint probability. 
But for a single light-curve made of real observations the model is unconstrained and even the random walk model may be inadequate \citep{graham:2014}.
As a consequence, parameters in equation~(\ref{eq:logaf}) may not be accurately estimated. 
Nonetheless, we can be fairly sure that for any two statistical tests probing the same null hypothesis on the same dataset, equation~(\ref{eq:logaf}) (or more accurately, a polynomial fit) still holds, and that the joint probability expressed by equation~(\ref{eq:jointprob}) cannot be easily calculated.
These results show that there is little point in performing these multitest probes of the same null hypothesis using the same data set because the results of one test are in principle predictable from the results of the other, and the joint probability is difficult to evaluate in practice.

\subsection{Double positive tests}

Things get even more complicated if the rejection of the multitest null hypothesis is not based on the multitest probabilities, but in the simultaneous rejection of the single tests involved, which I will call \emph{double positive tests} (DPT) practice. 
As a mutitesting procedure, DPT outcomes cannot be accurately described in terms of probabilities, and by construction, these outcomes are biased by the test that has the lower power. 
DPT is the actual methodology employed in some recent microvariability research \citep[e.g.,][]{joshi:2011,gaur:2012,hu:2013,chand:2014}.
For example, \citet{hu:2013} test variability in the BL~Lac object OI~090.4 using the \ctest and the \ftest, and they claim variability \emph{only} if both tests reject the null hypothesis at the significance level of $\alpha = 0.01$, or a \emph{dubious} event if only one of the tests reveals variability.
It is worth to note that in all the six dubious events reported in \citeauthor{hu:2013}, the test that fails to reject the null hypothesis is the \ctest, as expected from being the less powerful of the two tests involved.

\begin{figure}[t!]
\centering
Fig. in page 36
\epsscale{1.}
\caption{Comparison of the outcomes of the \ftest for the same quasar light-curve, but using two different comparison stars A and B.
This figure shows the results of 300 simulations.
The probabilities are highly correlated, as expected.
The regression fit hits the significance levels ($\alpha = 0.001$) intersection, as a result of both tests having the same power.
The expected number of Type~II errors for a single test is $300\alpha = 0.3$.
Thus, the most probable outcome of this multitest is that \emph{all} the DPT rejected variation events were true variations.}\label{fig:2ftests}
\end{figure}

\citet{joshi:2011}, \citetalias{gaur:2012}, and \citet{chand:2014} also implement DPT.
These authors apply the \ftest, but using two different stars as comparisons.
Again, evidence for variability is supported only if both \ftest simultaneously reject the null hypothesis. 
One problem with this procedure is that differences in brightness or variability in one of the stars may yield a low detection rate for the dimmer or variable star 
\citetext{scaled versions of the \ftest as in \citealt{howell:1988}, and \citealt{joshi:2011} have been proposed to handle the difference in brightness issue}.

It is clear that the dimmer star has larger magnitude errors than the brighter star.
These errors also produce a light-curve with larger variance for the dimmer star.
As the $F$ statistics is the ratio of the quasar and star light-curve variances, the larger the variance of the star, the lower power for detection of quasar variations.
Similarly, if one of the stars is itself variable, the variance of its light-curve will also be larger than for the steady star, and thus the power of the test performed with the variable comparison star will be lower.

In any case, if we were to compare the probabilities of the \ftest obtained with different comparison stars, we would obtain a relationship similar to that found when comparing \aov and \ftest, as shown in Figure~\ref{fig:2ftests}, with the only difference that in the double \ftest case the two probes would have the same power (assuming that both stars are non-variable and have the same brightness).
A single \ftest performed on these simulations yields about 60 variability detections, out of 300 events, while the double \ftest detects only 40 variations.
Given the number of simulated events (300) and the significance level of the single tests ($\alpha = 0.001$), we expected 0.3 false detections (Type~I errors) in our sample.
Therefore, we have considered 20 more events as non-variables, although they were almost certainly varying. 
As a result, we have increased dramatically the rate of Type~II errors.
Besides, the quantification of the joint probability of the $F$-DPT is as cumbersome as with any other multitest procedure.

\begin{figure*}[t!]
\centering
Figs. (a-b) in page 37, (c-d) in page 38
\caption{Single and two comparison stars versions of the \ftest. 
Panel (a): the histogram shows the distribution of the $F$ statistics for 3000 simulated differential light-curves of 35 observations each for a non-variable object compared through $F$-tests with a given star S$_a$; the thick line shows the theoretical PDF for the $F_{34,34}$ statistics scaled to the area of the histogram, and the vertical line indicates the location of the corresponding critical value for an upper tail test at the level of significance $\alpha = 0.01$ ($F_c = 2.2583$). 
The same data for the object was also tested against a second comparison star S$_b$ (not shown) with the same characteristics as S$_a$.
Panel (b): the histogram of the two comparison stars \ftest for the same object compared to the stacked data for stars S$_a$ and S$_b$ is shown along with the scaled theoretical $F_{34,68}$ PDF (thick line); the vertical line indicates the critical value for an upper tail test at $\alpha = 0.01$ ($F_c = 1.9452$). 
Panel (c): histogram similar to panel (a), but the object has been modeled with \rw microvariations; the thick line is the PDF of the $F_{34,34}$ statistics for the null hypothesis, scaled to the area of the histogram.
Panel (d): histogram similar to panel (b), but for the variable object shown in panel (c), and the scaled $F_{34,68}$ PDF (thick line).}\label{fig:fmulti}
\end{figure*}

\subsection{A power enhanced version of the \ftest}


Above in this section we have seen that multitesting produces outcomes that are difficult to characterize in terms of the level of significance for the combined tests.
Thus, combining different test procedures yields highly correlated probabilities, and if a DPT approach is chosen, the result is strongly biased by the test with the lower power.
However, in the CCD frames around a target quasar, there may be several comparison stars that, intuitively, we would like to include in the analysis to get an improved tool to detect microvaribility events.
Besides, the inclusion of several comparison stars in a test should reduce the possibility of fake microvariability detections that any peculiar single star light-curve might produce, a concern that was in the background of the DPT implementation.
Below a simple procedure is presented to expand the power of the \ftest using these field stars.

Basically, the strategy consists in increasing the number of the degrees of freedom in the denominator of the $F$-distribution of reference for the null hypothesis by stacking the  light-curves of the comparison stars.
Ideally, all the comparison stars and the quasar should be of equal brightness, but in practice their respective light-curves variances should be scaled by a term $\omega$ to compensate the larger photometric errors for dimmer objects \citep[e.g.][]{howell:1988,joshi:2011}.

Let us suppose that the light-curve of a given quasar has been observed $N_q$ times, along with a number of $k$ comparison stars. 
It is not necessary that the comparison stars have been observed the same number of times as the quasar or each other, although this is probably the case for CCD differential photometry.
For our purpose, let $N_j$ be the number of observations in the $j$-star light-curve.
For each star light-curve, we calculate the mean magnitude $\overline{m}_j$.
Then, for each observation $i$ of the $j$-star light-curve, we calculate a scaled square deviation:
\begin{equation}
    s_{j,i}^2 = \omega_j (m_{j,i} - \overline{m}_{j})^2,
\end{equation}
where $\omega_j$ is the term to scale the variance of the $j$-star to the level of the quasar $q$.
Stacking all the $s_{j,i}$ for all the observations and comparison stars, we can calculate the combined variance of the stars:
\begin{equation}
    s_c^2 = \frac{1}{(\sum_{j=1}^k N_j) - k} \sum_{j=1}^k \sum_{i=1}^{N_j} s_{j,i}^2.
\end{equation}
We will compare this combined variance with the quasar light-curve variance to get the $F$-statistics with $\nu_q = N_q -1$ degrees of freedom in the numerator, and $\nu_c = (\sum_{j=1}^k N_j) - k$ degrees of freedom in the denominator.
If the quasar and all the comparison stars have the same number of observations $N$ in their light-curves, the number of degrees of freedom can be expressed by $\nu_q = N -1$ and $\nu_c = k (N-1)$, respectively.

An increase in the number of degrees of freedom in the $F$-statistics, either in the numerator or in the denominator, yields also an increase in the power of the \ftest.
To increase the number of degrees of freedom in the numerator, we necessarily need more observations of the quasar; but we can multiply the number of degrees of freedom in the denominator by simply staking several field stars light-curves.
This result is discussed below using two comparison stars and both non-variable and variable quasar light-curves, and comparing the results with those obtained using a single comparison star.

Figure~\ref{fig:fmulti} shows scaled $F$-PDFs for the null hypothesis (solid lines), and histograms of the $F$-statistics values calculated for the single and multiple comparison stars versions of the \ftest.
Four sets of 3000 simulated light-curves, each one comprising 35 observations, were generated for two comparison stars S$_a$ and S$_b$, a non-variable quasar, and a \rw variable quasar.
The null hypotheses $F$-distributions of reference are $F_{34,34}$ for the single test, and $F_{34,68}$ for the two comparison stars \ftest, and their respective critical values at the $\alpha = 0.01$ level are $F=2.2583$ and $1.9452$.
Figure~\ref{fig:fmulti}a shows the results for the single comparison star version of the \ftest, with the $F$-statistics calculated as the ratio of the variances of non-variable quasar and star S$_a$ light-curves (results for star S$_b$ are similar, thus they are not shown).
Figure~\ref{fig:fmulti}b shows the results for the two comparison stars version of the \ftest, with the $F$-statistics calculated as the ratio of the variance of the same non-variable quasar light-curves and the combined variance of stars S$_a$ and S$_b$.
Figure~\ref{fig:fmulti}c-d are similar to Figure~\ref{fig:fmulti}a-b but for \rw variable quasar light-curves.


\begin{table}[t]
\begin{center}
\caption{Single and two comparison stars $F$-tests.\label{tab:fmulti}}
\begin{tabular}{lcc}
\tableline\tableline
Comparison    &    non-variable    &    Variable \\
\tableline
Star S$_a$    &    26              &    1876 \\
Star S$_a$    &    35              &    1863 \\
Stacked       &    34              &    2164 \\
\tableline
\end{tabular}
\end{center}
\end{table}


\begin{table*}[t]
\begin{center}
\caption{1ES\,2344+514: Summary of results at $\alpha=0.01$ presented in \citetalias{gaur:2012}.\label{tab:gaur}}
\begin{tabular}{lccccccc}
\tableline\tableline
                       &\multicolumn{3}{c}{$B$ band}&    &\multicolumn{3}{c}{$R$ band} \\
\cline{2-4}
\cline{6-8}
                              & Star 1  && Star 2   &         & Star 1  && Star 2   \\
\tableline
\cline{2-8}
Number of light-curves        &        &14&        &          &        &19 \\
\ctest                        &    0    &&    0    &          &    0    &&    0 \\
\ftest                        &    0    &&    3    &          &    1    &&    2 \\
\ftest double detections      &        &0&         &          &        &1 \\
\aov                          &    3    &&    2    &          &    9    &&    3 \\
\aov double detections        &        &2&         &          &        &3 \\
\tableline
\end{tabular}
\end{center}
\end{table*}

Table~\ref{tab:fmulti} summarizes the results of the different $F$-tests at the level of significance of $\alpha = 0.01$.
As the number of light-curve simulations is 3000 in each case, we expect about 30 Type~I error false detections for the non-variable source.
The results of the simulations for the non-variable source agree with this expectation for all the tests.
For the variable source, the single comparison star tests yielded about 1870 detections [$(62 \pm 1)$\%], while the two comparison stars \ftest was able to detect 2164 events [$(72 \pm 1)$\%].
Note that this power is similar to the results for \aov and Bartels test shown in Table~\ref{tab:testcomp}. 

The result of the multiple comparison stars \ftest presented here differs with the DPT and other multitesting procedures discussed above that include data from several comparison stars in the statistical analysis of light-curves.
Thus, the comparison star light-curves considered are used to perform only one \ftest, rather than several, highly correlated tests.
For this reason, the probability of the result is obtained directly from the test, rather than being indeterminate or replaced by a non-statistical quality criterium, as in the DPT.
Finally, the multiple comparison stars \ftest is a power enhance procedure, rather than a restrictive one like DPT; adding more comparison stars results in more power and thus more detections, while extending the concept of DPT to several comparison stars in a kind of multiple positive tests methodology would drop the number of detections dramatically.

The power enhancement produced by the two comparison stars \ftest is consistent with the expectations raised by the general methodology of stacking several comparison star light-curves discussed above.
In spite of the gain in power, the results are still hampered by the \ftest problem with the non-normal distribution of variable light-curves observations (see Section~\ref{sec:analytical}), and thus two comparison stars are needed to obtain results comparable with \aov and Bartels test using a single star.
However, if the number of bright and non-variable stars in the field around the quasar is large (i.e., one reference star for differential photometry and at least two comparison stars), the enhanced \ftest presented here may overcome the loss of power due to non-normality. 
Thus, this test is probably the most reliable and powerful procedure to detect microvariations in quasar light-cruves of those tests analyzed in this paper.
These results pose the question on how to implement correct multitesting procedures with other tests.


\section{Comparison with observations}\label{sec:observations}


The aim of this subsection is to compare the results of the simulations considered above with real observations obtained from AGN optical microvariability literature.
Comparing these results with real data is challenging.
For simulations, we are able to build a controlled model for variations (for example a random-walk of fixed drifts) and set the number of observations to produce different light-curves arising from an otherwise unique, ideal stochastic process. 
Therefore, we can test this process many times to estimate power and robustness for a given test, and for comparison with the results of other tests.
Of course, we can control neither the behavior of a real quasar, which can vary through different mechanisms, nor even the number of nights that we can dedicate to monitor a given source, which may be limited by atmospheric conditions, the duration of the research project, time allocation, and human resources.
As a result, the set of all observed light-curves will be more scarce and noisy than the simulated data.

For the purpose of comparing the results of simulations, the \citetalias{gaur:2012} paper is, to my knowledge, the only one that analyzes blazar light-curves using the \ctest, the \ftest, the \chisq test, \aov, and two comparison stars.
The \chisq test is not considered in the present paper, because it is seldom used in quasar microvariability studies and, as it was shown in \citetalias{diego:2010}, it is not a reliable test for comparing light-curves.
Using \citetalias{gaur:2012} results we will compare the power of the \ctest, \ftest and \aov, the relation between the results of the \ctest and \ftest, and some possible problems arising from DPT using two comparison stars. 

\citetalias{gaur:2012} monitored blazars 1ES\,1959+650 and 1ES\,2344+514 several nights through the years 2009 and 2010, with five telescopes located in India, Greece and Bulgaria.
Surprisingly for blazar objects, \citetalias{gaur:2012} do not claim any microvariability detection in either band ``as the $C$, $F$, \chisq and \aov results never showed significance levels above 99 per cent considering both stars.''
Although this statement is right for the \ctest and \chisq-test, for which there are not even a single detection, a closer inspection on \citetalias{gaur:2012} Tables 5 and 6 shows that in fact there are a few such double detections for the \ftest and \aov.
In the rest of this subsection, I will focus the discussion on the results for 1ES\,2344+514, mostly in the $B$ band.
Similar results can be obtained for the blazar 1ES\,1959+650 or in the $R$ band, but such in depth analysis of results presented in \citetalias{gaur:2012} is beyond the scope of this paper, where they are used for illustration purposes only.
The difference between the two comparison stars used in \citetalias{gaur:2012} is larger for 1ES\,2344+514 (stars C2 and C3, $\Delta R = 1.2$) than for 1ES\,1959+650 (stars 4 and 6, $\Delta R = 0.7$), and therefore it has larger effects on the derived light-curves for the former blazar, and makes the interpretation more straightforward.


Table~\ref{tab:gaur} summarizes the results obtained by \citetalias{gaur:2012} for the blazar 1ES\,2344+514. 
The first column indicates the topic, either the number of observed light-curves or the test type. 
The second to fourth columns show the results in the $B$ band for Star~1 (second column) and Star~2 (fourth column); the third column show information that is common to both stars. 
The fifth to seventh columns are like columns two to four, but for results in the $R$ band.
Three of the \aov double detections are significant even at $\alpha = 0.001$ (e.g., the $B$ light-curve for August 29, 2009 shown in \citetalias{gaur:2012} Figure~3 presents a variation of about 0.25\,mag detected by \aov, but not by the other tests).
An immediate result of a close inspection of Table~\ref{tab:gaur} is that, adding up the number of detections by test and band, results in zero detections for the \ctest in both $B$ and $R$ bands; 3 detections for the \ftest in both the $B$ and $R$ bands; and 5 and 12 detections for \aov in the $B$ and $R$ bands, respectively.
The comparative results for the \ctest and \ftest agree with those reported by \citet{hu:2013}, and \citet{joshi:2011} in the sense that microvariablity detections are more numerous for the \ftest than for the \ctest.
The comparison of the three test also agree with our expectations about the powers of the \ctest, \ftest, and \aov for the simulations presented here and in \citetalias{diego:2010}, viz. the power is lower for the \ctest, average for the \ftest, and higher for \aov.

\begin{figure}[t]
\centering
Fig. in page 39
\epsscale{1}
\caption{Logarithmic probabilities relationships between the \ctest and the \ftest for the blazar 1ES\,2344+514 results presented in \citetalias{gaur:2012}.
The dotted line shows the fit of an order three polynomial to enhance the relation between the probabilities.
}\label{fig:gaur_c_vs_F}
\end{figure}

\subsection{Comparison of the distribution of the $C$ and $F$ statistics}

In Figure~\ref{fig:moreregres}b we saw the relationship between the probabilities resulting from the \ctest and the \ftest for 300 \rw simulated light-curves.
All the probabilities for the \ctest in the simulations are below the empirical limit $\log(P) \lessapprox -0.3$ (i.e. $P < 0.5$), rather than $\log(P) < 0$ (i.e. $P < 1$), as we would expect for a honest test.
Figure~\ref{fig:gaur_c_vs_F} shows the relationship between the probabilities obtained for the \ctest and the \ftest from \citetalias{gaur:2012} data.
Despite the different scales in Figures~\ref{fig:moreregres}b and \ref{fig:gaur_c_vs_F}, it can be appreciated that the results obtained from observed light-curves agree with the results drawn from simulated data in the sense that there is a tight relationship between the $C$ and $F$ tests, and that the \ctest presents an empirical upper limit $\log(P) \lessapprox -0.3$, in accordance to the value obtained from the simulations. 

\begin{figure*}[t!]
\centering
Fig. (a) in page 40, (b) in page 41
\caption{PDFs and CDFs for the \ctest and \ftest.
Panel (a): The double of the normal PDF associated to the null hypothesis of the \ctest (solid line) and the $F$-PDF with $\nu_1 = nu_2 = 34$ degrees of freedom associated to the null hypothesis of the \ftest (dashed line).
Note that the $F$ values have been square root transformed to be represented as a function of the ratio of standard deviations rather than variances. 
Panel(b): The corresponding CDFs associated to the null hypothesis of the \ctest (solid line) and the \ftest (dashed line). 
Note the large difference between both CDF curves for small values of the $\sigma_1 / \sigma_2$ ratio.}\label{fig:cftests}
\end{figure*}

Figure~\ref{fig:cftests} presents a comparison between the \ctest and \ftest PDFs and CDFs (cumulative distribution functions) for the null hypothesis of non-variability; note that the areas below both PDF curves are normalized.
The distribution for the null hypothesis of the \ctest is postulated to be a positive normal random variable $z = \sigma_1 / \sigma_2$, with a PDF that is the double of the normal PDF to ensure normalization. 
With respect to the \ftest, the $F$-PDF and the $F$-CDF, both with $\nu_1 = \nu_2 = 34$ degrees of freedom, have been transformed in the sense that the statistics values are the standard deviation ratios rather than the variance ratios.

From Figure~\ref{fig:cftests} and some calculations we can understand why the empirical maximum of the $\log(P)$ probabilities for the \ctest is around -0.3 rather than 0.
For $\log(P) = -0.3$, both $P$ and CDF ($=1-P$) are approximately 0.5, and from the $C$-statistics we obtain this CDF corresponds to the quantile $\sigma_1 / \sigma_2 = 0.67$. In contrast, the $F$-CDF for the corresponding variances ratio $\sigma_1^2 / \sigma_2^2 = 0.45$ with $\nu_1 = \nu_2 = 34$ degrees of freedom is 0.01.
The large difference between the postulated $C$-CDF and the much more accurate $F$-CDF explains the scarcity of large $P$-values for the \ctest.
This is a serious problem because, by construction, the region of large $P$-values (or small CDF-values) for the \ctest corresponds to the mode of the $C$-distribution, i.e., the maximum of the PDF curve (see Figure~\ref{fig:cftests}a).
Similarly, the critical values for the significance level of $\alpha = 0.01$ are quite different: $\sigma_1 / \sigma_2 = 2.576$ for the \ctest, and $\sigma_1 / \sigma_2 = 1.503$ (or more properly, $\sigma_1^2 / \sigma_2^2 = 2.258$) for the \ftest with $\nu_1 = \nu_2 = 34$ degrees of freedom.

These differences between the $C$ and $F$ distributions increase with the number of observations included in the light-curve.
The mean and the standard deviation of the $F$ distribution are expressed respectively by:
\begin{equation}\label{eq:muF}
    \mu_F = \frac{\nu_d}{\nu_d - 2}, \text{ for } \nu_d > 2,
\end{equation}
and
\begin{equation}
    \sigma_F = \left[ \frac{2 \nu_2^2 (\nu_1 + \nu_2 -2)}{\nu_1 (\nu_2-2)^2 (\nu_2-4)} \right]^{1/2}, \text{ for } \nu_d > 4.
\end{equation}
This standard deviation becomes smaller for larger degrees of freedom, while the mean of the $F$ distribution basically does not change. 
As a consequence, the distribution of the $\sigma_1 / \sigma_2$ ratios also becomes narrower with increasing degrees of freedom and stable mean approaching asymptotically to 1.
Therefore, also the transformed $F$-PDF displayed in Figure~\ref{fig:cftests}a will turn narrower as the degrees of freedom get larger, while the $C$-distribution will not change.
Curiously, the \ctest, which is based on the normal distribution with infinite degrees of freedom, is more and more inaccurate as the number of observations increases.


\subsubsection{Multitesting with two comparison stars}

\citetalias{gaur:2012} provide the $F$ statistics and critical values $F_c$ for the blazar differential light-curves compared with two stars.
These $F$ values are given by:
\begin{equation}
    F_i = \frac{\sigma_{q,i}^2}{\sigma_{1,2}^2},
\end{equation}
with the same number $n-1$ of degrees of freedom in the numerator and in the denominator, and where $n$ is the number of observations in the light-curve. 
The subindex $i$ indicates the comparison star (1 or 2), $\sigma_{q,i}^2$ is the variance of the light-curve for the target source $q$ and star\,$i$, and $\sigma_{1,2}^2$ is the variance of the light-curve for stars\,1 and 2.
Fortunately, in this particular case the ratio:
\begin{equation}
    \frac{F_2}{F_1} = \frac{\sigma_{q,2}^2}{\sigma_{q,1}^2},
\end{equation}
is also a ratio of variances and thus it is $F$ distributed with $n-1$ degrees of freedom both in the numerator and in the denominator.
The $F$ distribution has a mean $\mu_F$ given by equation~(\ref{eq:muF})
Because $\mu_F \rightarrow 1$ very rapidly as $\nu_d \rightarrow \infty$, in the following paragraphs the value 1 is adopted instead of $\mu_F$.

\begin{figure}[t]
\centering
Fig. in page 42
\epsscale{1.}
\caption{PDFs for the ratios between the $F$-tests for the blazar 1ES\,2344+514 presented in \citetalias{gaur:2012}.
Three PDF curves are shown: dashed line for the data with the smaller number of observations (15), solid line for data with an average number of observations (43), and dotted line for data with the larger number of observations (85).
The statistic values are clearly biased towards values $F_2/F_1 > 1$, indicating differences between the statistics $F_2$ and $F_1$.
}\label{fig:gaur_PDF_vs_F}
\end{figure}

Star\,1 is the brightest, therefore we expect that the quasar light-curve for this star ($m_q - m_1$) may have a lower dispersion than the light-curve for the dimmer Star\,2, ($m_q - m_2$), and thus $F_2/F_1 > 1$.
That this effect may be detected or not will depend on the number of observations in each light-curve, the intrinsic quasar variability, the difference in brightness between the stars, and the relative contribution of the quasar and the stars to the light-curve dispersion either by genuine variations or shot noise.
For example, if the quasar variations are large, they dominate the scattering of the light-curves ($F_2/F_1 \simeq 1$). 
However, if the light-curve dispersion is dominated by shot noise, and the difference in brightness between the stars is large enough, relatively brighter Star\,1 will produce less differential light-curve scattering than the dimmer Star\,2 ($F_2/F_1 > 1$).
In the case of \citetalias{gaur:2012} blazars, variability does not dominate the light-curve statistics.


\begin{figure}[t]
\epsscale{1.}
\plotone{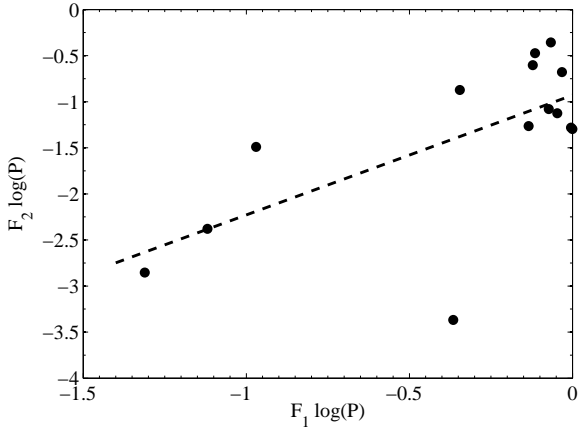}
\caption{Comparison of the outcomes of the \ftest for the blazar 1ES\,2344+514 light-curves in the $B$ band presented in \citetalias{gaur:2012}, using two different comparison stars A and B.
The logarithms of the probabilities for each light-curve are shown as filled circles.
The dashed line shows the linear regression for the logarithms of the probabilities.
}\label{fig:gaur_P2_vs_P1}
\end{figure}

Figure~\ref{fig:gaur_PDF_vs_F} shows the PDFs for the $F_2/F_1$ ratios in the $B$ band for the blazar 1ES\,2344+514 light-curves.
Taking into account that in the fair case that both stars yield the same variance in the blazar light-curves the ratios would be distributed around $F_2/F_1 \simeq 1$, the remarkable bias in the actual ratios shown in Figure~\ref{fig:gaur_PDF_vs_F} indicates that the light-curve variances of the stars are not the same.
This result is easily understood when considering that Star\,1 is substantially brighter (1.2\,mag) than Star\,2, and thus $F_2 > F_1$.
Therefore, tests performed using Star\,2 have lower power than those performed using Star\,1.

Finally, Figure~\ref{fig:gaur_P2_vs_P1} shows the relation between the probabilities obtained for the \ftest using both Star\,1 and Star\,2 as comparisons.
Despite the small number of observations and different scales, comparison with Figure~\ref{fig:2ftests} reveals similar trends in both simulations and real observations.
%
%
Comparison between the \ftest and \aov is not shown because the relationship is not as tight as for the \ftest and the \ctest, or two $F$ tests.
Therefore, the relation between the \ftest and \aov is somehow concealed by the small number of observations and scarce microvariability detections that would extend the graph ranges towards extreme low probabilities.


\section{Conclusions}\label{sec:conclusions}

This paper presented the analytical justification of the \aov and \ftest powers based on non-central $F$ distributions.
Predictive power analysis and inferred power obtained from simulations showed accurate agreement for \aov study of quasar light-curves. 
The \aov procedure is preferred when there is a limited number of suitable comparison stars in the quasar field for CCD differential photometry, but note that \aov requires oversampled light-curves which may limit its application. 

\begin{figure*}
\centering
\tikzstyle{decision} = [diamond, draw, 
    text width=6.5em, text badly centered, node distance=4cm, inner sep=0pt]
\tikzstyle{block} = [rectangle, draw, 
    text width=6em, text centered, node distance=3cm, rounded corners, 
    minimum height=4em]
\tikzstyle{line} = [draw, -latex']
\tikzstyle{cloud} = [draw, ellipse, 
    node distance=5cm, minimum height=2em]
    
\begin{tikzpicture}[node distance = 2cm, auto]
    \node [block, text width=10em] (init) {Microvariability tests};
    \node [decision, below of=init] (decide) {How many non-variable bright stars?};
    \node [block, below of=decide, left of=decide, node distance=3.5cm] (ftest) {Enhanced \ftest};
    \node [decision, below right of=decide, node distance=5cm] (onestar) {Oversampled light-curve?};   
    \node [block, below left of=onestar, node distance=4cm] (anova) {\aov};
    \node [block, below right of=onestar, node distance=4cm] (bartels) {Bartels test};
    \path [line] (init) -- (decide);
    \path [line] (decide) -| node [near start, left, above] {$\geq 3$} (ftest);
    \path [line] (decide) -| node [near start, above] {$<3$} (onestar);
    \path [line] (onestar) -| node [near start, above] {yes} (anova);
    \path [line] (onestar) -| node [near start, above] {no} (bartels);
\end{tikzpicture}
\vspace{0.5cm}
\caption{Decision flowchart for microvariability tests. If at least one reference and two comparison non-variable stars are available, the enhanced \ftest is preferred because of its power and reliability. When less than three bright stars are available, other procedures may be chosen. For oversampled light-curves, \aov is a powerful and robust test with internal error estimation. In other cases, Bartels test is also a powerful and robust nonparametric choice.}\label{fig:flowchart}
\end{figure*}
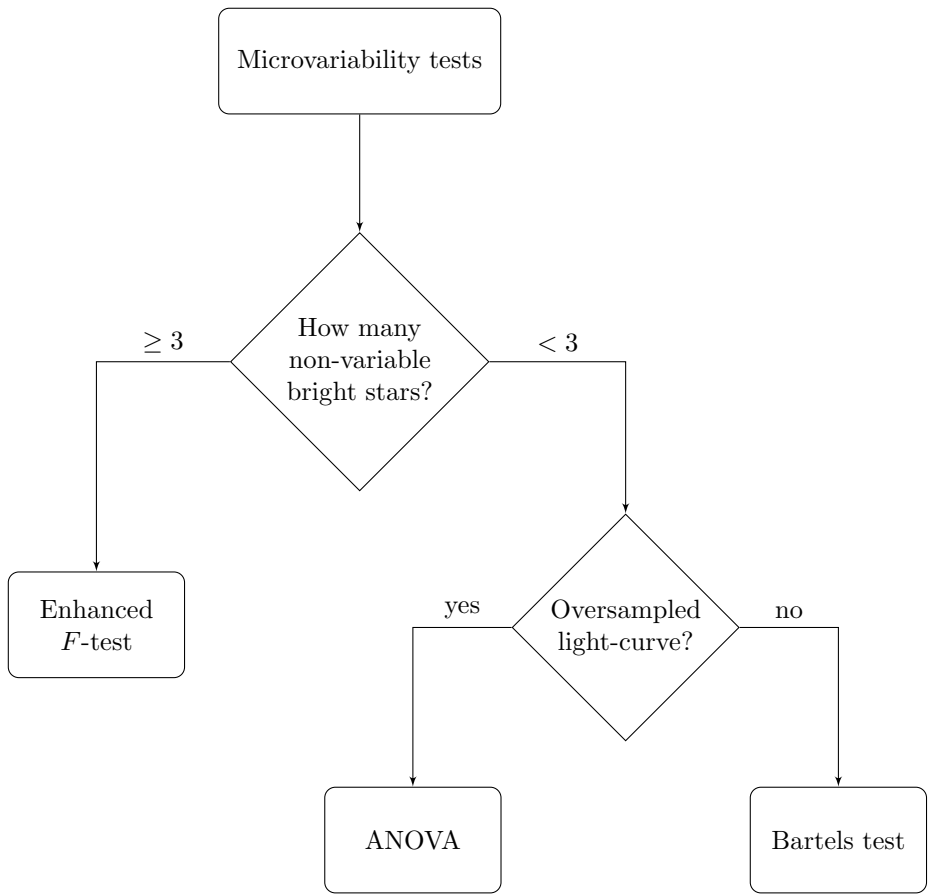

In contrast with \aov, \ftest for variances showed departures from the expected power values for non-Gaussian distributed data. 
Thus, \emph{post-hoc} power inferred from the simulations were significantly lower than the \emph{a priori} power prevision for the \ftest. 
This loss of power may be critical for variability detection when the quasar light-curve is compared with only one star, and in this case the researcher would rather try other procedures such as \aov or Bartels test.
However, a \ftest implementation that includes the light curves of two or more comparison stars has been presented in this paper. 
Including several comparison stars in the analysis also enhances the reliability of the test by diminishing possible odd effects caused by a single star.
This procedure overcomes the power limitations of the previous \ftest, and it is the preferred test if the quasar field contains several potential comparison stars.
The flowchart presented in Figure~\ref{fig:flowchart} summarizes the procedure to select a statistical test depending on the number of available bright stars and if the light-curve is oversampled or not.
	
The powers of the \ctest and of two nonparametric procedures, the Runs test and Bartels test, were also analyzed using simulations.
The \ctest showed an extremely low power, in accordance with the results stated in \citetalias{diego:2010}.
A close look at the $C$ and the $F$ distributions revealed large discrepancies that diminished the power of the \ctest.
The origin of these discrepancies resides in the \ctest wrong postulate about the akin to normal distribution for standard deviations ratios of quasars and comparison stars light-curves.
The \ctest should be avoided, especially taking into account that there are other, more rigorous ways to analyze the same datasets, such as the $F$ and Bartels tests. 

The nonparametric Bartels test showed an amazing capability to detect microvaribility in simulated light-curves, comparable to \aov.
Nonparametric tests are usually less sensitive than their parametric counterparts, particularly if the conditions of the parametric procedures are sufficiently met.
This is the case of the Runs test, that showed a rather low power in the simulations.
In contrast to \aov, the Bartels test requieres neither light-curve oversampling nor any special data gathering strategy.
Therefore, Bartels test becomes an interesting alternative to \aov and the \ftest, and it is worth to be probed in future research using real data, particularly if the \aov oversampling requirement cannot be met. 
	
The light-curve simulations also made evident the correlation between the results obtained with different tests, and that multitesting DPT procedures as implemented by other authors, always produce a remarkable and unjustified increase of Type~II errors.
It is surprising that such DPT procedures, that have less power, do not yield precise probabilities, and are more cumbersome and laborious than a simple single test, have gained their current levels of relevance and popularity. 
The results of DPTs to confirm variability detection are dominated by the test with the lowest power (either by the intrinsic nature of the tests involved, or the inherent noise of different comparison stars).
However, right procedures to combine several comparison field stars to enhance the power of the tests to detect microvariations exist, as it has been shown here in the case of multiple comparison stars enhanced \ftest.


Simulations presented in \citetalias{diego:2010} already demonstrated that \aov is a powerful test to detect microvariability in AGNs, a result that is confirmed from an analytical point of view in this paper. 
This paper has shown that the power of \aov and Bartels tests are very adequate to analyze microvariations in quasar light-curves when there is a limited number of comparison stars in the quasar field.
Apart from \aov, detection of microvariability events in quasars have also been reported using both $C$ and $F$ tests, or both together through DPT procedures. 
However, the inadequacy of the \ctest and of the DPT methodologies to yield reliable quantitative probabilities, and the loss of power of the \ftest applied to not normally distributed data, compromises the detection of genuine variations with these procedures.
Some of these problems have also been discussed comparing simulations and published microvariability results.
However, when several comparison stars are available in the quasar field, the power enhanced version of the \ftest presented in this paper may be the preferred statistical procedure to study microvariability in CCD quasar differential photometry.

\newpage 

This study is to my knowledge the first that applies nonparametric tests to the analysis of quasar light-curves, and the first that proposes a statistical rigorous methodology to include several comparison star light-curves to improve microvariability detection.
Besides, power analysis and data simulations have been successfully applied to investigate the possible outcomes of microvariability studies. 
Altogether, power analysis and simulations allow to compare the adequacy of different methodologies and to choose the best design of experiments to obtain the most from the researcher's effort.
The results presented here can improve future studies of microvariability, which in turn will help to understand the physical mechanisms that are responsable of this phenomenon.

\acknowledgments

This research has been supported by the UNAM-DGAPA-PAPIIT IN110013 Program, and the Canary Islands CIE: Tricontinental Atlantic Campus.
The author is thankful to the anonymous referee for the constructive suggestions.

\appendix

\section{Noncentral distributions}\label{app:ncd}

In hypothesis testing, central distributions characterize the behavior of a test statistics if the null hypothesis is true. 
But if the null hypothesis is false, the statistics that describes the data is a noncentral distribution. 
The parameter that defines the noncentral distribution is the noncentrality parameter $\lambda$. 
When $\lambda = 0$, the null hypothesis is true and the noncentral distribution is identical to the ordinary, central distribution \citep{murphy:2009}.


\subsection{Noncentral \chisq distribution}


Let $(X_1, X_2, \ldots X_k)$ be $k$ independent normally distributed random variables with means $\mu_i$ and variances $\sigma_i^2$, then: %
\begin{equation}
\chisq = \sum_{i=1}^k \frac{X_i^2}{\sigma_i^2},
\end{equation}
is a noncentral \chisq-distributed random variable. The noncentral \chisq distribution is characterized by two parameters, namely $k$ which is the number of degrees of freedom, and $\lambda$ usually called \emph{noncentrality} parameter:
\begin{equation}
\lambda = \sum_{i=1}^k \frac{\mu_i^2}{\sigma_i^2},
\end{equation}

\subsection{Noncentral $F$ distribution: test of equality of variances}\label{ssec:ftest}

Let $X$ be a noncentral \chisq random variable with noncentrality parameter $\lambda_{\chi^2}$ and $\nu_1$ degrees of freedom, and $Y$ a \chisq random variable with $\nu_2$ degrees of freedom. %
If $X$ and $Y$ are statistically independent, then: %
\begin{equation}\label{eq:F}
F = \frac{X_1^2 / \nu_1}{Y_2^2 / \nu_2}
\end{equation}
is a noncentral $F$-distributed random variable characterized by a noncentral parameter $\lambda$ given by: 
\begin{equation}
\lambda = \sigma^2 / \varepsilon^2,
\end{equation}
where $\sigma^2$ and $\varepsilon^2$ are the variances to be compared in the \ftest.

The equality of variances test requires the data in the two samples with variances $\sigma^2$ and $\varepsilon^2$ to be normally distributed.
This is not the usual case for quasar variability (see Section~\ref{sec:analytical} in the main text).
I have used a step function to simulate an easy to analyze quasar variations. Let us calculate $\lambda$ and the analytical power of the \ftest in this simplified variability example.
For this purpose, we define the \emph{effect size} $r$ as a measure of the strength of a variation in a set of quasar observations:
\begin{equation}
r = \frac{n}{N} \frac{\sigma_q^2}{\varepsilon^2},
\end{equation}
where $n = 5$ is the number of observations that vary in the step function of amplitude $\sigma_q = 0.04$ (true variations without photometric error term), $N = 35$ is the total number of observations, and $\varepsilon = 0.01$ the photometric error, resulting in $r = 2.286$.
Ideally, the error $\varepsilon$ should coincide with the standard deviation of the comparison star light-curve (if the star and the quasar have the same brightness).

To calculate the \ftest power, we need the scattering due to variability to be normally distributed across the $N$ light-curve observations rather than concentrated in $n$ data points.
Therefore, we need to calculate the variance $\sigma_r$ for this distribution such that it produces the same effect size as the step function light-curve.
Such variance is given by:
\begin{equation}
\sigma_r^2 = \varepsilon^2 \, (1+r),
\end{equation}
which yields a value of $\sigma_r^2 = 3.286 \times 10^{-4}$.
Now we can calculate the noncentrality parameter:
\begin{equation}
\lambda = \frac{\sigma_r^2}{\varepsilon^2},
\end{equation}
that is, $\lambda = 3.286$.

Power can be computed in R by:
\begin{displaymath}
    \mathrm{pf(}\lambda * F_\alpha^{\nu_1,\nu_2},\nu_1,\nu_2),
\end{displaymath}
where $F_\alpha^{\nu_1,\nu_2}$ is the $F$ statistics critical value corresponding to a level of significance $\alpha$ and degrees of freedom $\nu_1$ and $\nu_2$. %
In our case with $\lambda = 3.286$, $\nu_1 = \nu_2 = 34$, and  choosing $\alpha = 0.001$, this critical value is $F_{0.001}^{34,34} = 0.335$, and we obtain the test power as the probability that the measured $F$ statistics is $P(F\geq F_{0.001}^{34,34}) = 61\%$.

\subsection{Noncentral $F$ distribution: one way \aov}

For one-way \aov effect size is measured by $f$ \citep[e.g.][]{kabacoff:2011}:
\begin{equation}
    f = \sqrt{\frac{\sum_{i=1}^k p_i (\mu_i - \mu)^2}{\sigma^2}}.
\end{equation}
where the proportion $p_i = n_i /N$, with $n_i$ the number of observations in group $i$ and $N$ the total number of observations; the means $\mu_i$ for each group $i$, the grand (total) mean $\mu$, and the error variance or variance within groups 
$\sigma^2$. %
The \aov noncentrality parameter $\lambda$ is expressed by:
\begin{equation}
    \lambda = f^2 \times N.
\end{equation}

Adapting the example in \S\ref{ssec:ftest} to the \aov procedure, we now plan to achieve $N = 35$ observations, with an error $\varepsilon = 0.01$\,mag, but now the observations are carried out in $k=7$ groups of $n_i=n=5$ observations each, and with the quasar varying during one of these groups of observations (identified as the $j$ group), with an amplitude of $\Delta m_j = 0.04$. %
Thus, for the 6 $i$ groups that do not present variations, $\mu_i - \mu = 0$, and for the the data in the varying group $j$, $\mu_j - \mu = 0.04$. %
Then, the effect size is $f = 1.5119$, and the noncentrality parameter $\lambda = 80$, and the number of degrees of freedom of the noncentral $F$ distribution is $\nu_g = k-1 =6$ for groups and $\nu_r = N - k = 28$ for residuals.

Power for \aov can be calculated through the noncentrality parameter $\lambda$ as for the \ftest; in R it is possible to calculate it using commands `pf' or `power.anova.test' of the default loaded \emph{stats} package, or the command `pwr.anova.test' of the \emph{pwr} package. %
For example, for $k$ groups of $n$ observations each, and $f$ the effect size, and for a sigma level $\alpha$, the \aov power can be computed by:
\begin{displaymath}
    \mathrm{pf}(\lambda * F_\alpha^{\nu_g,\nu_r},\nu_g,\nu_r),
\end{displaymath}
or:
\begin{displaymath}
    \mathrm{pwr.anova.test}(k,n,f,\mathrm{sig.level=}\;\alpha).
\end{displaymath}

\section{Nonparametric Tests}


\subsection{Runs Test for Detecting Non-randomness}

The Runs test is used to examine wether a sequence of $n$ data occurred in random order (independently) or not.
There are several ways to define runs depending on the characteristics of the original data (e.g. head and tails for coin tosses, even and odd for counts, above and below the mean or the median for discrete and continuous data), but the final sequence produced must be dichotomous. 
In this paper, a run of length $l$ is a sequence of $l$ adjacent photometrical values all of them either above (coded +) or below (coded -) the mean of the light-curve.
A small number of runs indicates a tendency for large and small values to cluster, while a large number of runs indicates a tendency to oscillate.

Under the null hypothesis (random order), the number of runs $m$ in a sequence of $n$ observations is a random variable.
Let be $n_+$ the number of observations above the mean, and $n_-$ the number of observations below the mean ($n = n_+ + n_-$).
Then, $m$ is approximately normal distributed with mean and variance given by:
\begin{align}
    \mu & = \frac{2 n_+ n_-}{n} +1, \\
    \sigma^2 & = \frac{2 n_+ n_- (2 n_+ n_- - n)}{n^2 (n-1)}.
\end{align}

If both $n_+$ and $n_-$ are larger than 12, the test statistics may be approximated by a normal distribution with quantiles given by $z = (m - \mu) / \sigma$;
otherwise, exact solutions based on the number of ways of distributing $n$ observations into $m$ runs are preferred \citep[chapter 3 and Appendix Table~D]{gibbons:2003}.

\subsubsection{Test conditions}

The Runs test is a non-parametric statistical test, thus its reliability is not constrained to data that complies with a particular distribution.
Moreover, there is no assumption about the probabilities associated with positive and negative elements, and even the critical value for the dichotomous classification is arbitrary (the mean, the median, or any relevant value for the researcher).
Though, the data must be either dichotomous as collected, or coded into a dichotomous sequence depending if the observations is above or below some fixed quantity.
\citet{mogull:1994} demonstrated that the test cannot signal departures from randomness with run lengths of two.

The Runs test can be two-sided if the alternative hypothesis is randomness, or one sided if the alternative hypothesis is either the presence of a trend (left tail) or oscillations (right tail).

\subsubsection{Application for microvariability}

Runs test is easily implemented, and thus it is supported by most general purpose statistical software. In R code, the `runs.test' procedure is implemented in the \emph{tseries} package.

Using the Runs test to study light-curve microvariability implies that we assume that a single variation can be monitored several times, yielding a run of large length.
Therefore, the appropriate procedure is performing a low tail test (large lengths imply fewer runs).


\subsection{Bartels Test}


The Bartels test is also known as the \emph{rank version of von Newmann ratio test for randomness} \citep{bartels:1982}. 
As in the case of the Runs test, the Bartels test is also used to examine whether a sequence of $n$ data occurred in random order or not.
The test is based on the sum of squares of the rank differences between contiguous elements of a time sequence.
Let $R_i$ be the rank of the $i$th observation in a sequence of $n$ observations.
For large samples ($n > 10$), the test statistics is:
\begin{equation}
    \mathrm{RVN} = \frac{\sum_{i=1}^{n-1} (R_i - R_{i+1})^2}{\sum_{i=1}^{n} (R_i - \overline{R})^2},
\end{equation}
where $\overline{R} = (n+1)/2$ is the rank mean.
This RVN statistics is asymptotically normally distributed, with mean $\mu$ and variance $\sigma^2$ given by:
\begin{align}\label{eq:Bartels}
    \mu & = 2, \\
    \sigma^2 & = \frac{4(n- 2)(5 n^2- 2 n- 9)}{5 n (n+ 1) (n - 1)^2}.\label{eq:BartelsStats}
\end{align}

Data trends in the time sequence will produce small values of the RVN statistics.
Therefore the rejection region to test randomness against data trends (the alternative hypothesis considered in this paper), is small RVN values.
To be used in microvariability studies, we must assume that each microvariability event will be monitored several times, as in the Runs test.
Similarly, oscillations will produce large values of the RVN statistics, a possibility not considered in this work.

\subsubsection{Modifications to R bartels.test\{lawstat\} function.} 

An inspection of the R function bartels.test, included in the \emph{lawstat} package, showed that it uses a standard deviation that is derived from  the variance approximation $\sigma^2 \approx 4/n$, rather than the exact variance value given in equation~\ref{eq:BartelsStats}.
Because the number of elements in the simulations range from 15 to 35, some of them close to the limit of $n>10$ for the Normal approach, the approximated variance was changed by its exact value.
This change, negligible for a single test, enhances the accuracy of the results for a large number of simulations.

%
%

\bibliographystyle{apj}

\end{document}